\documentclass[%
 reprint,
superscriptaddress,
nofootinbib,
 amsmath,amssymb,
 aps,
]{revtex4-2}
\usepackage{graphicx}% Include figure files
\usepackage{dcolumn}% Align table columns on decimal point
\usepackage{bm}% bold math
\usepackage[version=4]{mhchem} %chemical elements
\usepackage{subfig}
\usepackage{float}
\usepackage{caption}
\usepackage{multirow}
\usepackage[colorlinks,
            linkcolor=red,
            anchorcolor=blue,
            citecolor=blue
           ]{hyperref}% add hypertext capabilities
\graphicspath{ {figure/} }
\begin{document}

\title{A rigorous solution to the superluminal issue in the diffusion equation}
%\title{superluminal diffusion of charged particles}
% Force line breaks with \\

\author{Xing-Jian Lv}
\email{lvxj@ihep.ac.cn}
 \affiliation{%
 Key Laboratory of Particle Astrophysics, Institute of High Energy Physics, Chinese Academy of Sciences, Beijing 100049, China}
\affiliation{
 University of Chinese Academy of Sciences, Beijing 100049, China 
}%
 \author{Xiao-Jun Bi}
 \email{bixj@ihep.ac.cn}
\affiliation{%
 Key Laboratory of Particle Astrophysics, Institute of High Energy Physics, Chinese Academy of Sciences, Beijing 100049, China}
\affiliation{
 University of Chinese Academy of Sciences, Beijing 100049, China 
}%
\author{Kun Fang}
\email{fangkun@ihep.ac.cn}
\affiliation{%
 Key Laboratory of Particle Astrophysics, Institute of High Energy Physics, Chinese Academy of Sciences, Beijing 100049, China}
 \author{Peng-Fei Yin}
\email{yinpf@ihep.ac.cn}
\affiliation{%
 Key Laboratory of Particle Astrophysics, Institute of High Energy Physics, Chinese Academy of Sciences, Beijing 100049, China}
\author{Meng-Jie Zhao}
\email{zhaomj@ihep.ac.cn}
 \affiliation{%
 Key Laboratory of Particle Astrophysics, Institute of High Energy Physics, Chinese Academy of Sciences, Beijing 100049, China}
\affiliation{
China Center of Advanced Science and Technology, Beijing 100190, China 
}%

%\affiliation{
%\affiliation{%


\date{\today}% It is always \today, today,

\begin{abstract}
Superluminal propagation is an intrinsic problem in the diffusion equation and has not been effectively addressed for a long time. In this work, a rigorous solution to this issue is obtained under the assumption that particles undergo a random flight process, where they move isotropically at a constant speed while experiencing random scatterings. We validate this solution by comparing it with comprehensive simulations of the random flight process and find that it significantly deviates from the solution derived from the J\"{u}ttner propagator. This solution is broadly applicable to various diffusion phenomena, such as cosmic-ray propagation. We emphasize that our rigorous solution is particularly crucial in scenarios involving burst-like particle injection, where previous phenomenological approaches to the superluminal diffusion problem may not yield accurate results.
%Specifically, in scenarios of continuous injection, such as the diffusion of electrons and positrons within pulsar halos, our results confirm previous findings based on phenomenological approaches. However, for the case of burst-like particle injection, our rigorous solution significantly deviates from previous phenomenological models.
\end{abstract}
\maketitle

%\tableofcontents

\section{\label{sec:intro}INTRODUCTION}
It is well established that the diffusion equation~\cite{beckerTheoryHeat1967, HANGGI1982207}
\begin{equation}
\frac{\partial}{\partial t} \rho(x, t)=D \frac{\partial^2}{\partial x^2} \rho(x, t), \quad t \geq 0, \quad D>0\label{enq: diff eqn}
\end{equation}
is fundamentally incompatible with the principles of special relativity. The inconsistency is apparent when considering the propagator of Eq.~(\ref{enq: diff eqn}), which is given by
\begin{equation}
p\left(x, t \mid x_0\right)=\frac{H(t)}{(4 \pi D t)^{1 / 2}} \exp \left[-\frac{\left(x-x_0\right)^2}{4 D t}\right]\;,
\label{eqn: Green Gaussian}
\end{equation}
where $H(t)$ denotes the Heaviside step function. This propagator corresponds to the solution of Eq.~(\ref{eqn: Green Gaussian}) for the initial condition of $\rho(x, t=0)=\delta(x-x_0)$. As indicated by Eq.~(\ref{eqn: Green Gaussian}), for any $t>0$, there exists a nonzero probability that the particle can be located at distances $|x-x_0| > ct$, where $c$ is the speed of light. Henceforth, we refer to this issue as superluminal diffusion.

The superluminal diffusion problem has garnered significant attention over the past century (see Ref.~\cite{Dunkel:2006kn} for review and references); yet, a definitive solution has remained elusive. While the telegraph equation~\cite{10.1093/qjmam/4.2.129, monin1955diffusion} provides an exact description for the isotropic walk of a particle with constant speed in one spatial dimension, where the random free path follows an exponential distribution. It has been demonstrated~\cite{uchaikinTelegraphEquationRandom2000} that in spatial dimensions $d\geq2$ -- which are highly relevant for numerous applications -- the simple diffusion offers a better approximation to the exact result than the more intricate telegraph equation.

The diffusion equation also plays a central role in astroparticle physics, as the propagation of charged particles through magnetic fields, such as the propagation of cosmic rays (CR), follows a diffusive process~\cite{1990acrbookB}. However, the issue of superluminal diffusion frequently arises in several contexts, such as the propagation of ultra-high energy CRs~\cite{Aloisio:2004jda, Berezinsky:2005fa, Aloisio:2008tx}, and the diffusion of electrons and positrons~\cite{Recchia:2021kty} within newly discovered pulasr halos~\cite{HAWC:2017kbo, LHAASO:2021crt, Fang:2022fof}.

Given the absence of a definitive solution to the superluminal diffusion problem, the astroparticle physics community often resorts to somewhat \textit{ad hoc} methods. For instance, in Ref.~\cite{Recchia:2021kty}, the authors manually combine the CR distributions of ballistic and diffusive motion, incorporating smoothing by hand. Meanwhile, Ref.~\cite{Bao:2021hey} employs the so-called generalized J\"{u}ttner propagator initially proposed by Ref.~\cite{Aloisio:2008tx}. However, this propagator is not derived from first principles; instead, it is an educated guess based on the relativistic generalization of the Maxwell-Boltzmann distribution. Consequently, in both approaches, although the limiting cases for $t\to0$ and $t\to\infty$ are correctly reproduced, no rigorous physical justification is provided for the intermediate regime.

In this work, we provide a \textit{rigorous} correction to the superluminal issue in the diffusion equation. We begin with a microscopic description of particles moving at a constant velocity while undergoing random isotropic changes in direction, and then derive the resulting propagator semi-analytically. Our findings are further validated through simulations. Our results indicate that the generalized J\"{u}ttner propagator does not accurately capture the intermediate regime of diffusion for particles moving at a constant speed. 
%However, in the case of continuously injecting sources, the time integration ensures that the final CR distribution is reasonably well reproduced by the generalized J\"{u}ttner propagator. 

The outline of this paper is as follows: we first revisit the treatment of the superluminal diffusion problem using the generalized J\"{u}ttner propagator Section~\ref{sec: previous}. The details of our approach are then presented in Section~\ref{sec:method}. Our results, along with a comparison to those obtained using the generalized J\"{u}ttner propagator, are discussed in Section~\ref{sec:results}. We further explore the application to pulasr halos in Section~\ref{sec:tevhalo}. Finally, the conclusions are provided in Section~\ref{sec:conclusion}

\section{\label{sec: previous}J\"{u}ttner propagator: a previous effort to the problem}
The generalized J\"{u}ttner propagator was put forward by~\cite{Aloisio:2008tx} (hereafter ABG10), where the authors utilized the fact that the propagator of the diffusion equation can be obtained from the Maxwell-Boltzmann speed distribution for particles of mass $m$ in a thermal equilibrium gas at temperature $T$, by making the substitution $v \to x$ and $2kT/m \to \lambda^2$. Given that the Maxwell-Boltzmann distribution has a relativistic extension -- known as the Maxwell-J\"{u}ttner distribution, which was originally derived by J\"{u}ttner~\cite{1911AnP339856J} and later modified by subsequent studies to more accurately represent the correct relativistic equilibrium distribution~\cite{Dunkel:2006nk}:
\begin{equation}
f_{\mathrm{M}-\mathrm{J}}(v)=\frac{\gamma^4}{4 \pi c^3} \frac{\mu}{K_1(\mu)} \exp (-\gamma \mu)\;,\label{eqn: mj}
\end{equation}
where $\gamma=1/\sqrt{1-v^2/c^2}$ represents the Lorentz factor, $\mu=mc^2/kT$, and $K_1$ is the first order
modified Bessel function of the second kind. ABG10 adopted Eq.~(\ref{eqn: mj}) as the basis and applied the same replacements $v \to x$, $2kT/m \to \lambda^2$ and $c\to ct$ treating it as the propagator for the relativistic diffusion problem, to derive the generalized J\"{u}ttner propagator:
\begin{equation}
\begin{aligned}
P_{\text {J\"{u}ttner}}&(E, r, t)=  \frac{1}{4 \pi(c t)^3} \frac{H[1-\xi(r, t)]}{\left[1-\xi^2(r, t)\right]^2} \\
& \times \frac{\kappa(E, t)}{K_1[\kappa(E, t)]} \exp \left[-\frac{\kappa(E, t)}{\sqrt{1-\xi^2(r, t)}}\right],\label{eqn: juttner prop}
\end{aligned}
\end{equation}
where $\xi=r / ct$ and $\kappa=2(ct / \lambda)^2$. ABG10 further demonstrated that Eq.~(\ref{eqn: juttner prop}) is a plausible approximation, as it correctly reproduces ballistic motion and pure diffusion in the appropriate limits. However, this remains an educated guess without a firm physical basis, especially considering that the two phenomena -- diffusion of particles and the velocity distribution in thermal equilibrium -- are fundamentally different in nature.

\section{\label{sec:method}Solution to the superluminal diffusion problem}
\begin{figure}[htbp]
\includegraphics[width=0.4\textwidth]{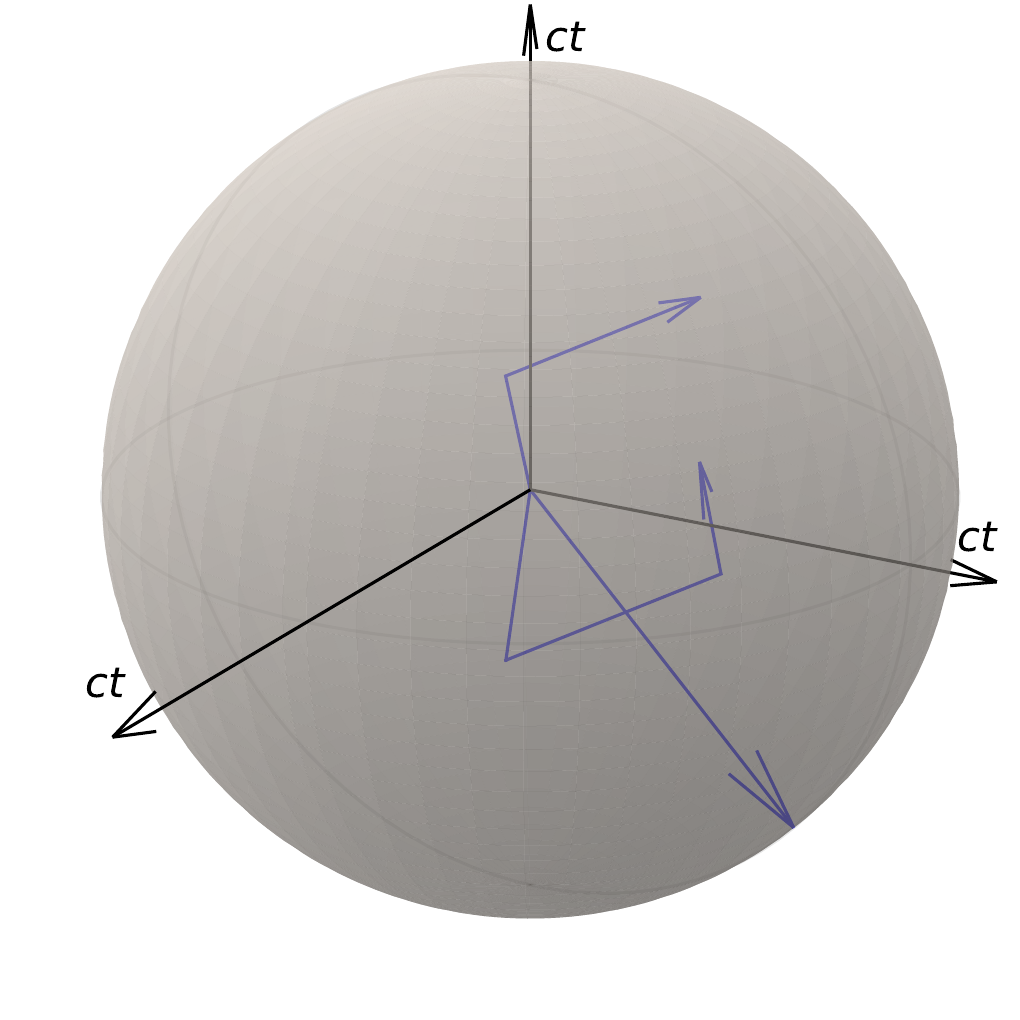}\\
\captionsetup{justification=raggedright}
\caption{Sample paths of the random flight in three spatial dimensions. The sphere represents the ballistic front, which consists of particles moving without any scattering. \label{fig:schematic}}
\end{figure}

%To avoid relying on hand-waving arguments by analogy, we must begin with a microscopic picture of particle diffusion. 
We begin with a microscopic picture of particle diffusion. Similar to ABG10, our goal is to directly compute the probability density function (PDF) of a relativistically acceptable diffusion model\footnote{We do not attempt to seek Lorentz or Poincar\'{e} invariant spatial diffusion processes, as diffusion theory is intended to offer a simplified phenomenological description of the complex stochastic motion of a particle within a background medium (in our case, the interstellar medium). Consequently, a preferred frame exists, corresponding to the rest frame of the substrate (cf. Ref.~\cite{Dunkel:2006kn}).}, rather than attempting to relativize the diffusion equation, as is done with the Schrödinger equation in quantum mechanics.

We begin by considering the ``simple random walk" model~\cite{lawlerRandomWalkModern2010}, wherein both time and space are discrete, and space is constrained to one dimension. This model serves as a foundation for understanding why the classical diffusion approach permits superluminal motion. A particle moves along the $x$-axis, starting from $x=0$ at $t=0$. During each discrete time interval of duration $\tau=1$, from $t=i$ to $t=i+1$ for $i=0,1,2,\dots$, the particle moves with velocity $v=\pm1$, with each direction having an equal probability of $1/2$. Consequently, the particle takes a step $h_i=\pm1$, with the steps being independent of one another. The position of the particle at time $t$, denoted as $x_t$, is given by $x_t = \sum_{i=0}^{t} h_i$. The expected value of $x_t$ is $\langle x_t \rangle = 0$, and the variance of $x_t$ is $\text{Var}(x_t) = \langle (x_t - \langle x_t \rangle)^2 \rangle = t$.

According to the central limit theorem in probability theory, for large values of $t$, the probability density of $x_t$ asymptotically approaches a normal distribution with a mean of 0 and a variance of $t$. This probability density, denoted as $\rho(x, t)$, can be expressed as
\begin{equation}
\rho(x, t) = \frac{e^{-x^2 / 2 t}}{(2 \pi t)^{1 / 2}}, \quad|x|=\mathcal{O}(\sqrt{t})\;, \label{eqn:diff1d}
\end{equation}
which satisfies the diffusion equation with a diffusion coefficient $D=1/2$ with the initial condition $\delta(x)$. The problem of Eq.~(\ref{eqn:diff1d}) is evident: for any $t > 0$, there is a nonzero probability density that the particle can be found at any position $x$, regardless of how large it is. This result contradicts the fact that $x_t$ has a maximum value of $|x_t| = t$, which is reached when all steps are taken in the same direction, implying that $\rho(x, t) = 0$ for $|x| > t$.

%It is important to recognize that Eq.~(\ref{eqn:diff1d}) is valid only for $|x| = \mathcal{O}(\sqrt{t})$. On the other hand, the condition $\rho(x, t) = 0$ applies when $|x| > t$. Therefore, these two results are applicable within different ranges of $x$. 
The problem of superluminal diffusion arises when Eq.~(\ref{eqn:diff1d}) is used outside its applicable domain, i.e., $|x| = \mathcal{O}(\sqrt{t})$. This issue can be straightforwardly resolved through a combinatorial idea. The exact solution of $\rho(x, t) = 0$, without yielding to any approximation, is provided by the binomial distribution:
\begin{equation}
    \rho(x, t) = \frac{1}{2^t} \binom{t}{(x+t)/2}\;.
\end{equation}

For the astrophysical scenario of interest, it is necessary to generalize the ``simple random walk" model to encompass continuous space and time, as well as higher dimensions. This generalization naturally leads to the so-called ``random flight" process~\cite{hughesRandomWalksRandom1995} (also known as the "persistent random walk"~\cite{WEISS2002381}), and we cite the definition of random flight provided by \cite{skogseidStatisticalMechanicsRandom2011} as follows:\\
\textit{A particle begins at the origin of the coordinate system and moves in a straight line with a constant speed $v$ in a random direction. The particle changes its direction at random times governed by a Poisson process with rate parameter $\lambda_0$. Each change of direction occurs instantaneously at a specific point, and the speed $v$ remains constant throughout the motion between direction changes. At each change, the new direction is chosen from an isotropic distribution, meaning all directions are equally likely.}

The sample paths of the moving particle are straight lines with sharp turns, forming polygonal shapes composed of randomly oriented segments of varying lengths. A schematic representation of this is shown in Figure~\ref{fig:schematic}, alone with the ballistic
front, which consists of particles moving without any
scattering. The mean free path is $l = v/\lambda_0$, which correspond to a diffusion coeffecient of $D_0 = lv/3 =  v^2/(3\lambda_0)$ in three spatial dimensions. Next, we approach this problem by solving it semi-analytically in three dimensions and then use simulations to verify the results.

\begin{figure*}[htbp]
    \begin{center}
        \subfloat{\includegraphics[width=0.31\textwidth]{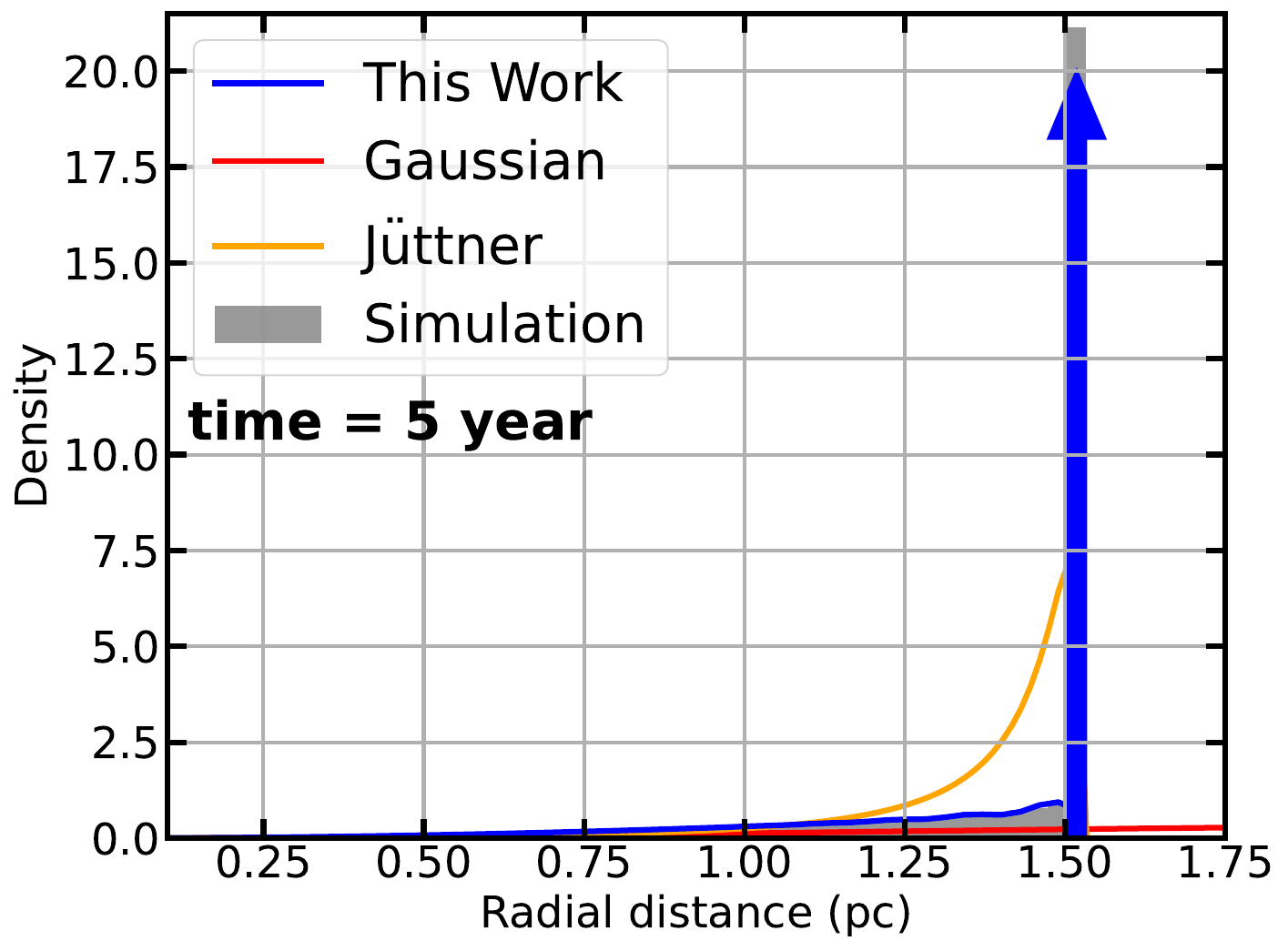}} \hskip 0.02\textwidth
        \subfloat{\includegraphics[width=0.31\textwidth]{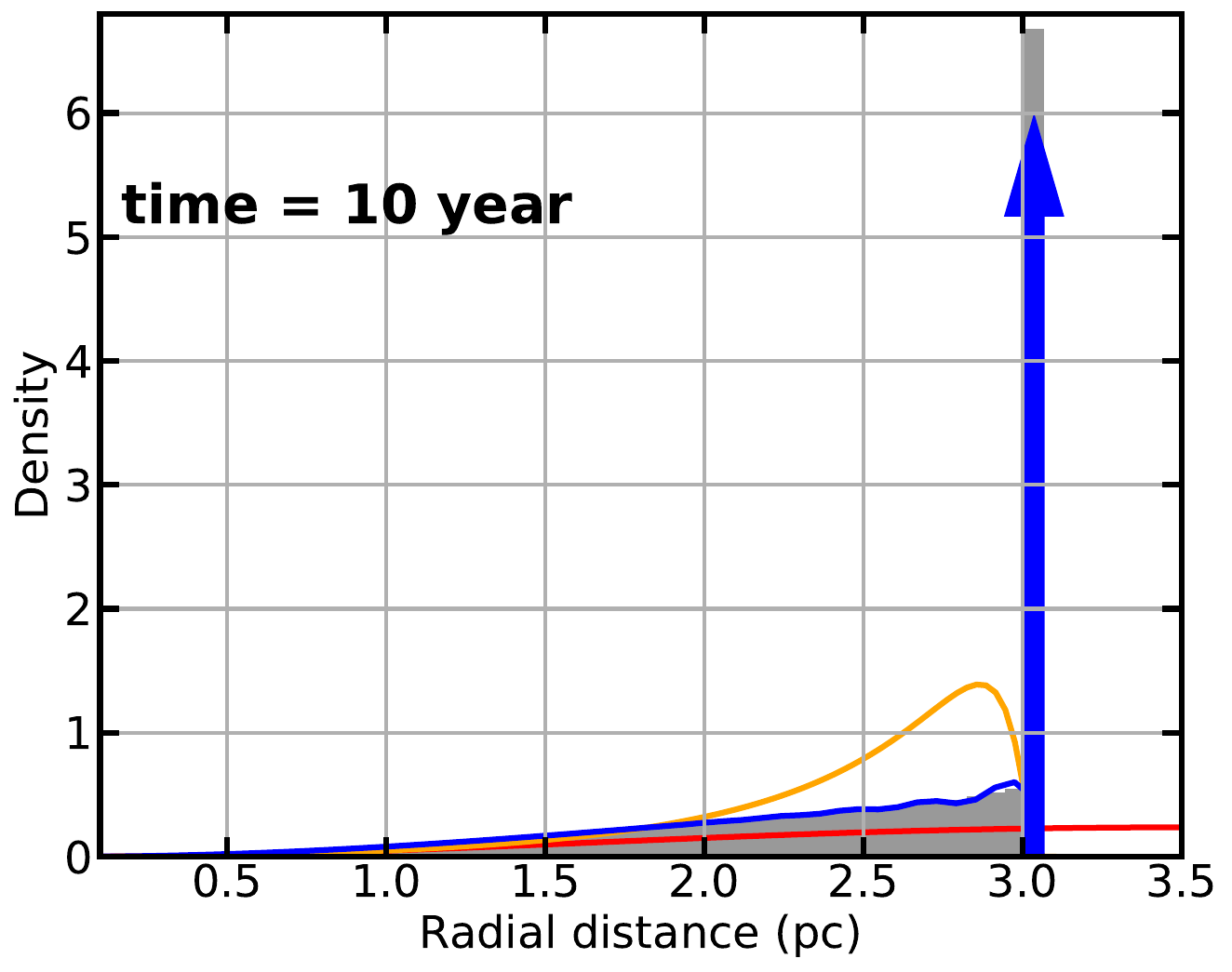}}
        \hskip 0.02\textwidth
        \subfloat{\includegraphics[width=0.31\textwidth]{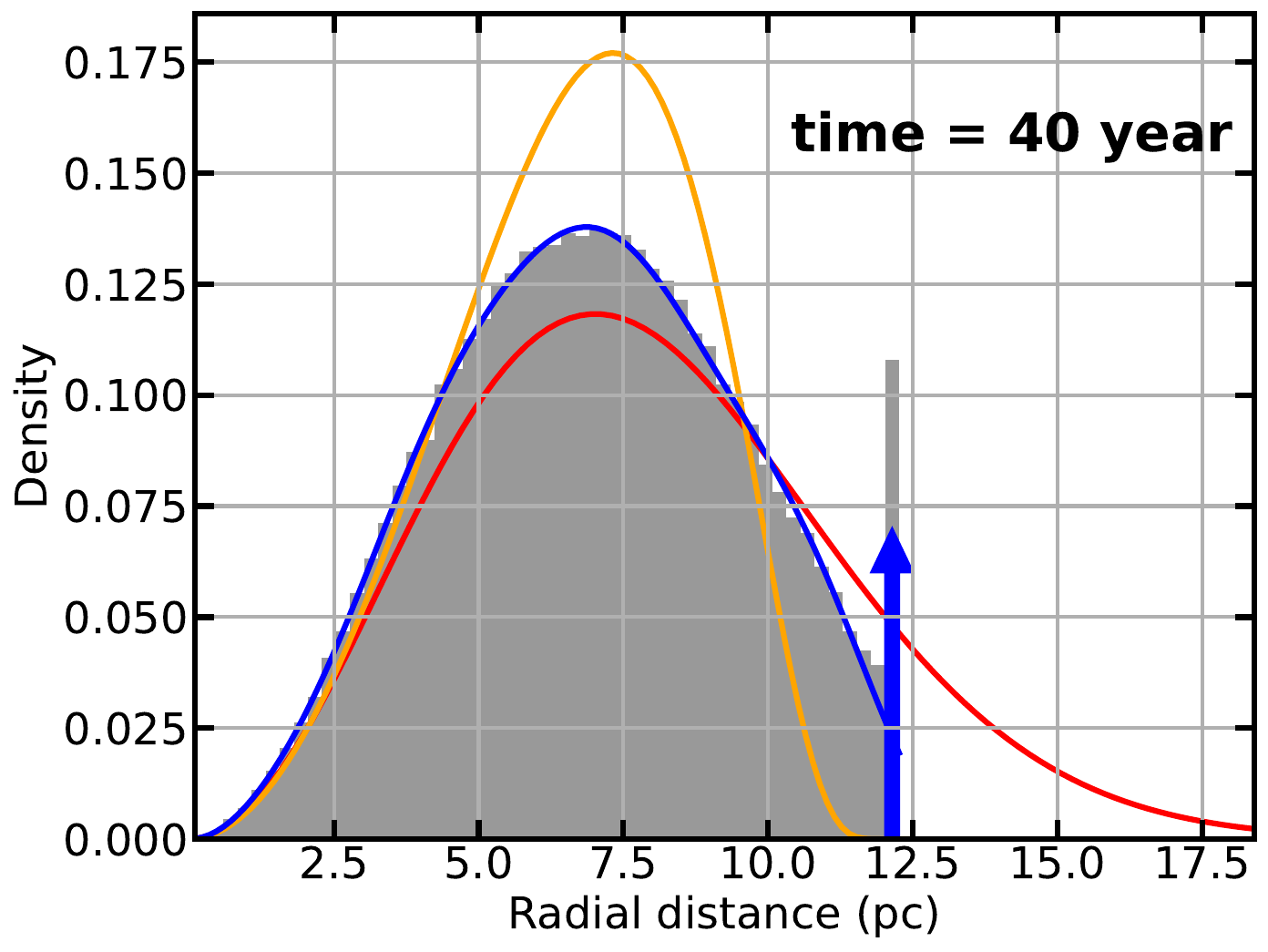}}\\
        \subfloat{\includegraphics[width=0.31\textwidth]{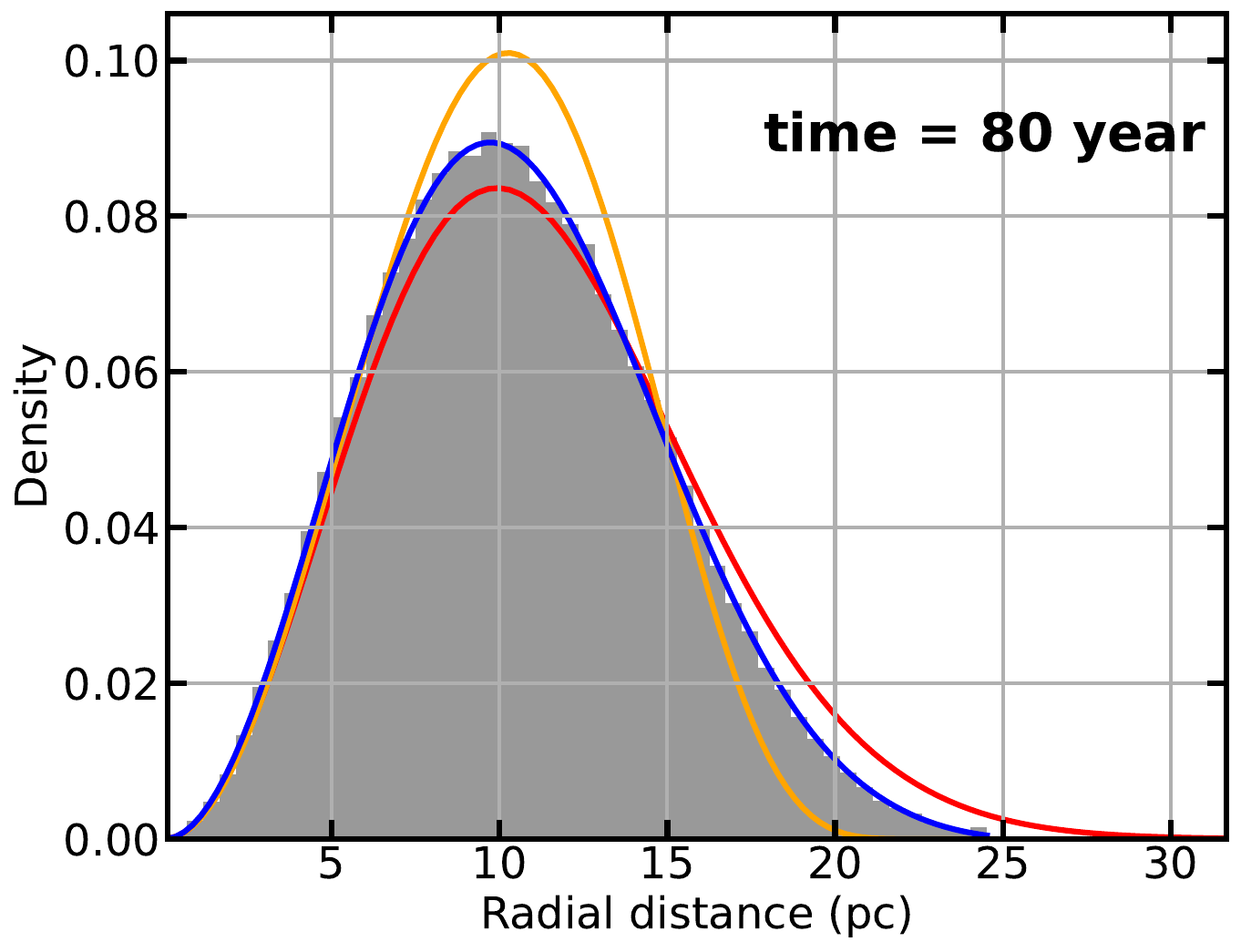}} \hskip 0.02\textwidth
        \subfloat{\includegraphics[width=0.31\textwidth]{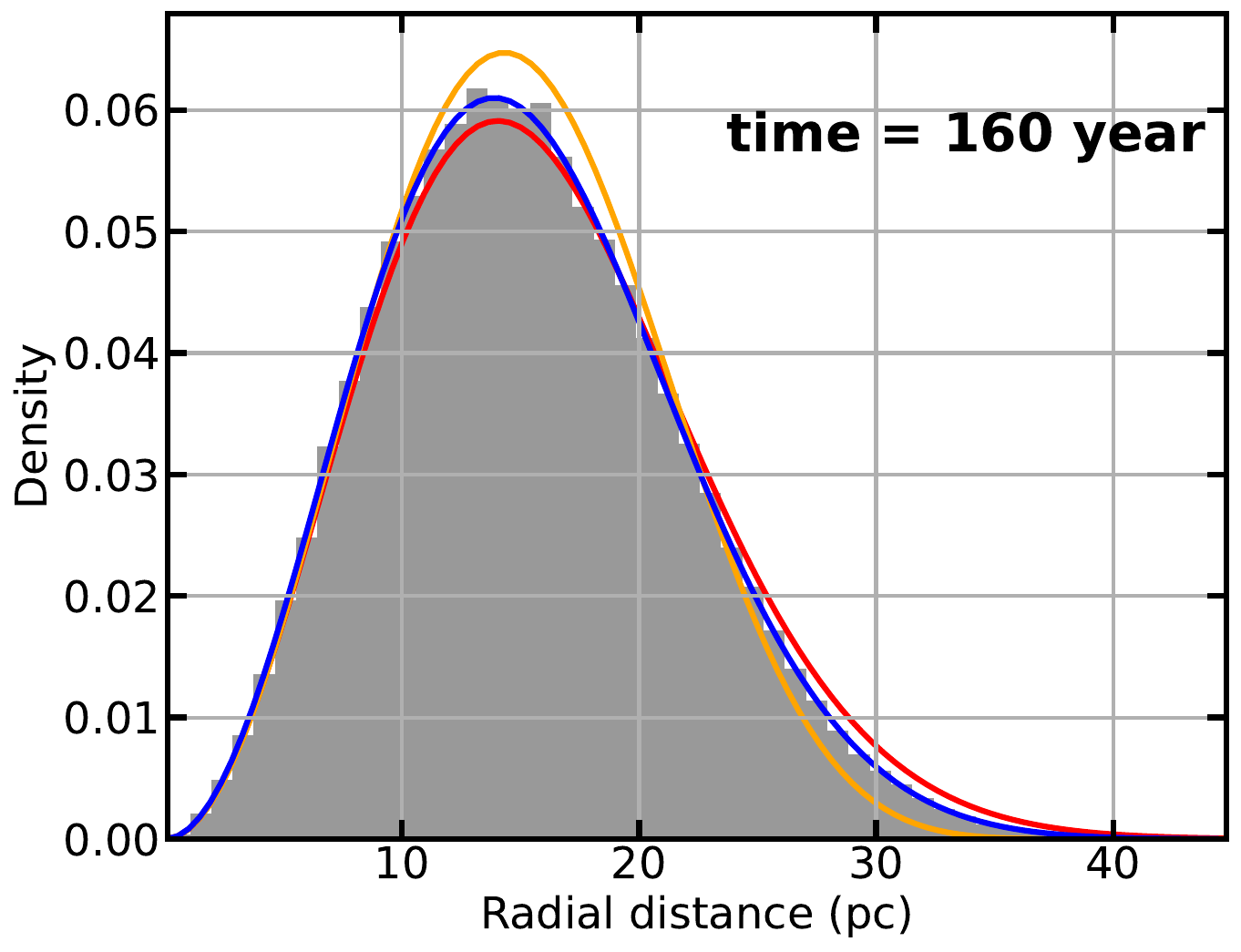}}
        \hskip 0.02\textwidth
        \subfloat{\includegraphics[width=0.31\textwidth]{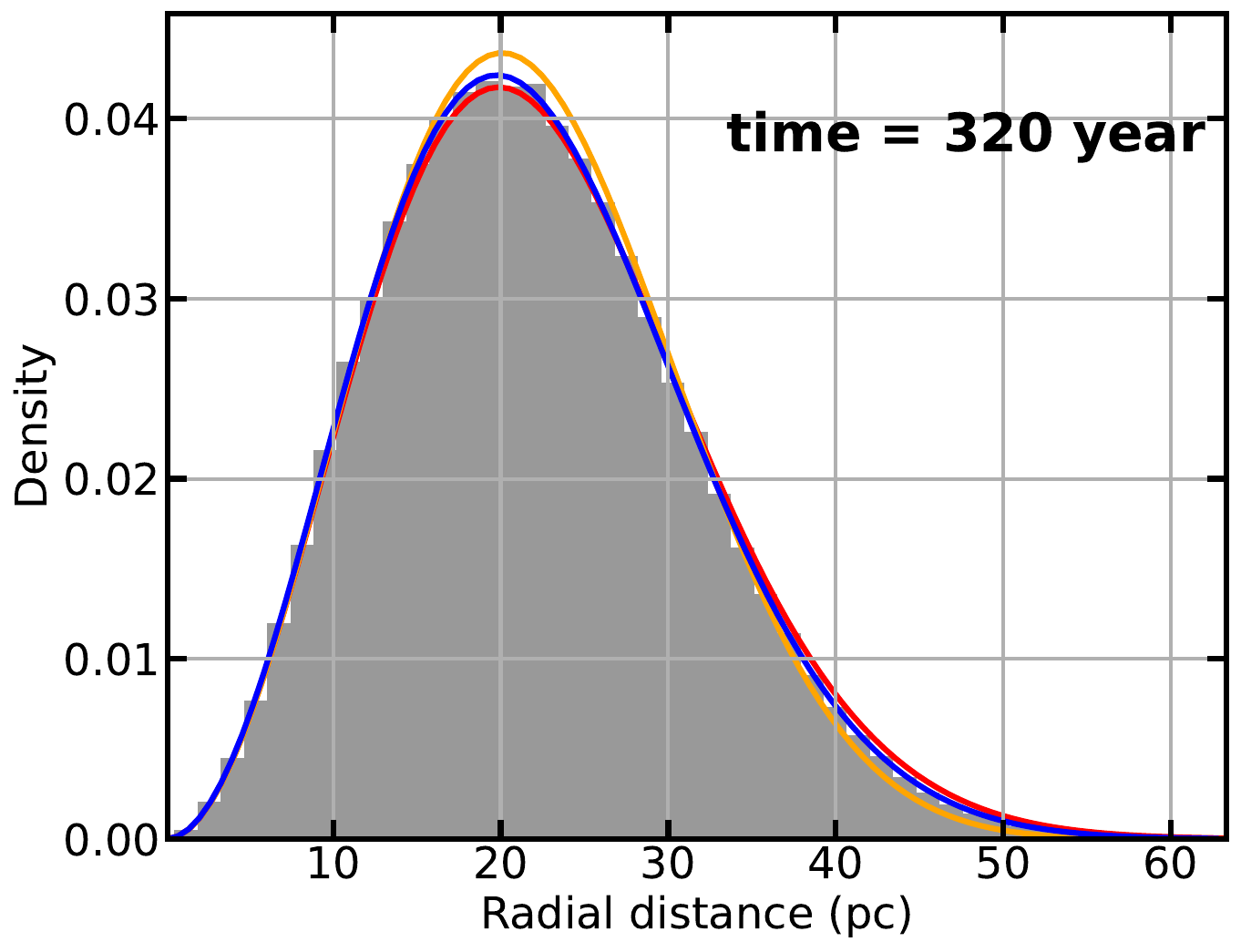}}
    \end{center}
    \captionsetup{justification=raggedright}
    \caption{The PDF as a function of radial distance $r$ for different times $t$, using the Gaussian propagator, the generalized Jüttner propagator, and the approach developed in this work, alongside a comparison with simulation results. The arrows represent $\delta$-peaks, corresponding to contributions from non-scattered particles. The mean time between scatterings is set to $\tau=10$ year.}\label{fig:green.}
\end{figure*}

\subsection{analytical solution}
We define: $\eta(\vec{x}, t) \equiv e^{-\lambda_0 t}\delta(r-vt)/V(r, t)$, where $V(r, t) = (2\pi^{n/2}r^{n-1})/\Gamma(n/2)$ is the surface area of an n-dimensional sphere, and $r=|\vec{x}|$. The term $e^{-\lambda_0 t}$ gives the probability of a particle moving without scattering, while the factor $v(r, t)$ describes the surface area of an n-dimensional sphere. Thus,  $\eta(\vec{x}, t)$ represents the PDF of particles located at the ballistic front. The function $\rho(\vec{x}, t)$ is then the solution to the integral equation~\cite{skogseidStatisticalMechanicsRandom2011}:
\begin{equation}
\begin{aligned}
\rho(\vec{x}, t)=&\eta(\vec{x}, t)+\lambda_0\left(\eta \otimes \rho\right)(\vec{x}, t)\\
=&\eta(\vec{x}, t)+\lambda_0\int_0^t d t^{\prime} \int d \vec{x}^{\prime} \eta\left(\vec{x}^{\prime}, t^{\prime}\right) \rho\left(\vec{x}-\vec{x}^{\prime}, t-t^{\prime}\right),\label{eqn: master}
\end{aligned}
\end{equation}
where $\otimes $ denotes space-time convolution.
This equation expresses that, at a given $(\vec{x}, t)$, a particle either has scattered or not. In the case where it has scattered, it did so for the first time at some earlier point $(\vec{x}', t')$, where the process began anew, generating $\rho$ with its origin at $(\vec{x}', t')$.

To undo the space convolution in the above equation, we apply the Fourier transform, denoted by $\widetilde{\phantom{a}}$, and to undo the temporal convolution, we employ the Laplace transform, denoted by $\hat{\phantom{a}}$. The Fourier-Laplace transform is thus defined as follows:
\begin{equation}
\hat{\tilde{f}}(\vec{k}, s)=\int_{\mathbb{R}^n} \mathrm{~d}^n \vec{r} \int_0^{\infty} f(\vec{r}, t) \exp [-\mathrm{i} \vec{k} \cdot \vec{r} - s t] \mathrm{d} t\;. \label{eqn: fourier-laplace trans}
\end{equation}
Given the spherical symmetry implied by the definitions of the random flight, the resulting transforms will depend on $\vec{k}$ solely through its magnitude, $k = |\vec{k}|$.
Applying Eq.~(\ref{eqn: fourier-laplace trans}) to Eq.~(\ref{eqn: master}) results in a straightforward algebraic equation:
\begin{equation}
\hat{\tilde{\rho}}=\hat{\tilde{\eta}}+\lambda_0 \hat{\tilde{\eta}} \hat{\tilde{\rho}}\;,
\end{equation}
from which $\hat{\tilde{\rho}}$ can be solved as:
\begin{equation}
\hat{\tilde{\rho}}=\frac{\hat{\tilde{\eta}}}{1-\lambda_0\hat{\tilde{\eta}}}\;.\;\label{eqn: solution}
\end{equation}

Substituting the Fourier-Laplace transform of $\eta(\vec{x}, t)$ in three dimensions~\cite{stadjeExactProbabilityDistributions1989}:
\begin{equation}
\hat{\tilde{\eta}}_3(\vec{k}, s)=\frac{\tan ^{-1}\left(\frac{v k}{\lambda_0+s}\right)}{vk}\;,
\end{equation}
we obtain the Fourier-Laplace transform of $\rho_3(\vec{x}, t)$~\cite{claesRandomWalkPersistence1987a, MASOLIVER1993469}:
\begin{equation}
\hat{\tilde{\rho}}_3(\vec{k}, s)=\frac{\tan ^{-1}\left(\frac{v k}{\lambda_0+s}\right) }{v k-\lambda_0 \tan ^{-1}\left(\frac{v k}{\lambda_0+s}\right)}\;.\label{eqn: 3D solution}
\end{equation}

The problem is completely solved by inverting the Fourier-Laplace transform of $\hat{\tilde{\rho}}$. However, this inversion can only be performed analytically in 1, 2, and 4 dimensions~\cite{skogseidStatisticalMechanicsRandom2011}, utilizing a semigroup-type property of Bessel functions. The corresponding solutions are presented in Appendix~\ref{sec:eqns}. Unfortunately, for the three-dimensional case, which is of primary interest, the inversion of the transformed equation $\hat{\tilde{\rho}}_3(\vec{k}, s)$ remains unresolved.

Therefore, to obtain a practical solution, we need to numerically invert Eq.~(\ref{eqn: 3D solution}). Directly taking the numerical Fourier-Laplace inverse of this equation may lead to instability due to the presence of the delta function, which arises from the contributions of the unscattered particles at the ballistic front. Given that we know the form of $\rho_3(\vec{x}, t)$ must be:
\begin{equation}
    \rho_3(\vec{x}, t) = e^{-\lambda_0 t}\frac{\delta(r-v t)}{4 \pi r^2} + \rho^{\text{mul}}_3(\vec{x}, t)\;\label{eqn: 3D solution mod},
\end{equation}
where $\rho^{\text{mul}}_3$ is the PDF corresponding to one or more scatter(s), we first subtract the Fourier-Laplace inverse of the no-scatter term:
\begin{equation}
    \mathcal{FL}(e^{-\lambda_0 t}\frac{\delta(r-v t)}{4 \pi r^2}) = \frac{\tan ^{-1}\left(\frac{v k}{\lambda_0+s}\right)}{vk}\;.
\end{equation}
Next, we numerically compute the Fourier-Laplace inverse of the remaining part:
\begin{equation}
\begin{aligned}
        \hat{\tilde{\rho}}^{\text{mul}}_3(\vec{k}, s) &=\hat{\tilde{\rho}}_3(\vec{k}, s) - \frac{\tan ^{-1}\left(\frac{v k}{\lambda_0+s}\right)}{vk}
        \\
        &=\frac{\tan ^{-1}\left(\frac{v k}{\lambda_0+s}\right)^2}{vk(vk-\lambda_0\tan ^{-1}\frac{v k}{\lambda_0+s})}\;.
\end{aligned}
\end{equation}
This approach allows us to isolate and handle the unscattered contribution separately, ensuring a more stable numerical inversion process. We employ the Stehfest algorithm~\cite{10.1145/361953.361969} for the numerical inverse Laplace transform, implemented in the \texttt{invertlaplace} function from the Python package \texttt{mpmath}\footnote{\url{https://mpmath.org/doc/current/}}~\cite{mpmath}, with the precision set to \texttt{mpmath.mp.dps = 25} for enhanced accuracy. For the inverse Fourier transform, due to the spherical symmetry of the random flight process, we only need to carry out a single integral given by the radially symmetric n-dimensional inverse Fourier transform:
\begin{equation}
\begin{aligned}
    f(r) &= \frac{1}{(2\pi)^n} \int_{\mathbb{R}^n} \hat{f}(\vec{k}) e^{i \vec{k} \cdot \vec{r}} \, d^n\vec{k} \\
         &= \frac{1}{(2\pi)^{\frac{n}{2}} r^{\frac{n}{2}-1}} \int_0^\infty \hat{f}(k) k^{\frac{n}{2}} J_{\frac{n}{2}-1}(kr) \, k \, dk\;,
\end{aligned}
\end{equation}
where $J_\nu$ is the Bessel function of the first kind of order $\nu$. 
This is a Hankel transform, and its numerical implementation is performed using the \texttt{SymmetricFourierTransform} function, with accuracy parameters set to \texttt{N=3200} and \texttt{h=0.001}, from the Python package \texttt{hankel}\footnote{\url{https://zenodo.org/records/3235680}}~\cite{Murray2019}.

\subsection{simulations}
To assess the accuracy of the numerical Fourier-Laplace inversion, we simulate a random flight process involving $2 \times 10^5$ particles in three dimensions. Each particle begins at the origin, selects a random direction isotropically, and moves along this direction at a constant velocity $v$. The duration of each movement is drawn from an exponential distribution with rate parameter $\lambda_0$. After completing each displacement, the process repeats for a total of 1000 steps. For each particle, we record both the duration of each move, $t_i$, and the corresponding position, $\boldsymbol{r}_i$. To determine the position of a particle at a given time $T$, we identify the time interval such that $t_{i-1} < T < t_i$. The particle position at time $T$ is then interpolated between $\boldsymbol{r}_{i-1}$ and $\boldsymbol{r}_i$ using uniform linear motion, given by the expression:
\begin{equation}
    \boldsymbol{t}_T=\boldsymbol{r}_{i-1} + \frac{T-t_{i-1}}{t_i-t_{i-1}}\times(\boldsymbol{r}_{i}-\boldsymbol{r}_{i-1})\;.
\end{equation}

Separate simulations for random flight processes with varying 
$\lambda$ values are unnecessary due to the existence of the following scaling symmetry: $\{r, t\} \to \{ \zeta r, t/\zeta \}$, where $\zeta$ is an arbitrary none-zero parameter.
%The particle distribution at time $t$ for a random flight process with scattering rate $\lambda$ is equivalent to that of a process with rate $\lambda_0$, after applying the rescaling transformations: $t\to\lambda_0t/\lambda$ and $r\to\lambda/\lambda_0$.

\section{\label{sec:results}Results \& Discussion}
The main results of this work are presented in Fig.~\ref{fig:green.}, where we plot the PDF as a function of radial distance $r$ for different times $t$, using the Gaussian propagator, the generalized J\"{u}ttner propagator, and the approach developed in this work, alongside a comparison with simulation results. We take the mean time between scatterings to be \( \tau = 10 \) years, which corresponds to a diffusion coefficient of \( D_0 \sim 10^{29} \, \mathrm{cm}^2\mathrm{s}^{-1} \). To represent the $\delta$-peaks, which arise from the contributions of non-scattered particles, we use vertical arrows with widths corresponding to the bin sizes employed for plotting the simulation histogram and heights normalized such that the area of each arrow matches the contribution of the respective $\delta$-peak.

\begin{figure}[htbp]
\includegraphics[width=0.48\textwidth]{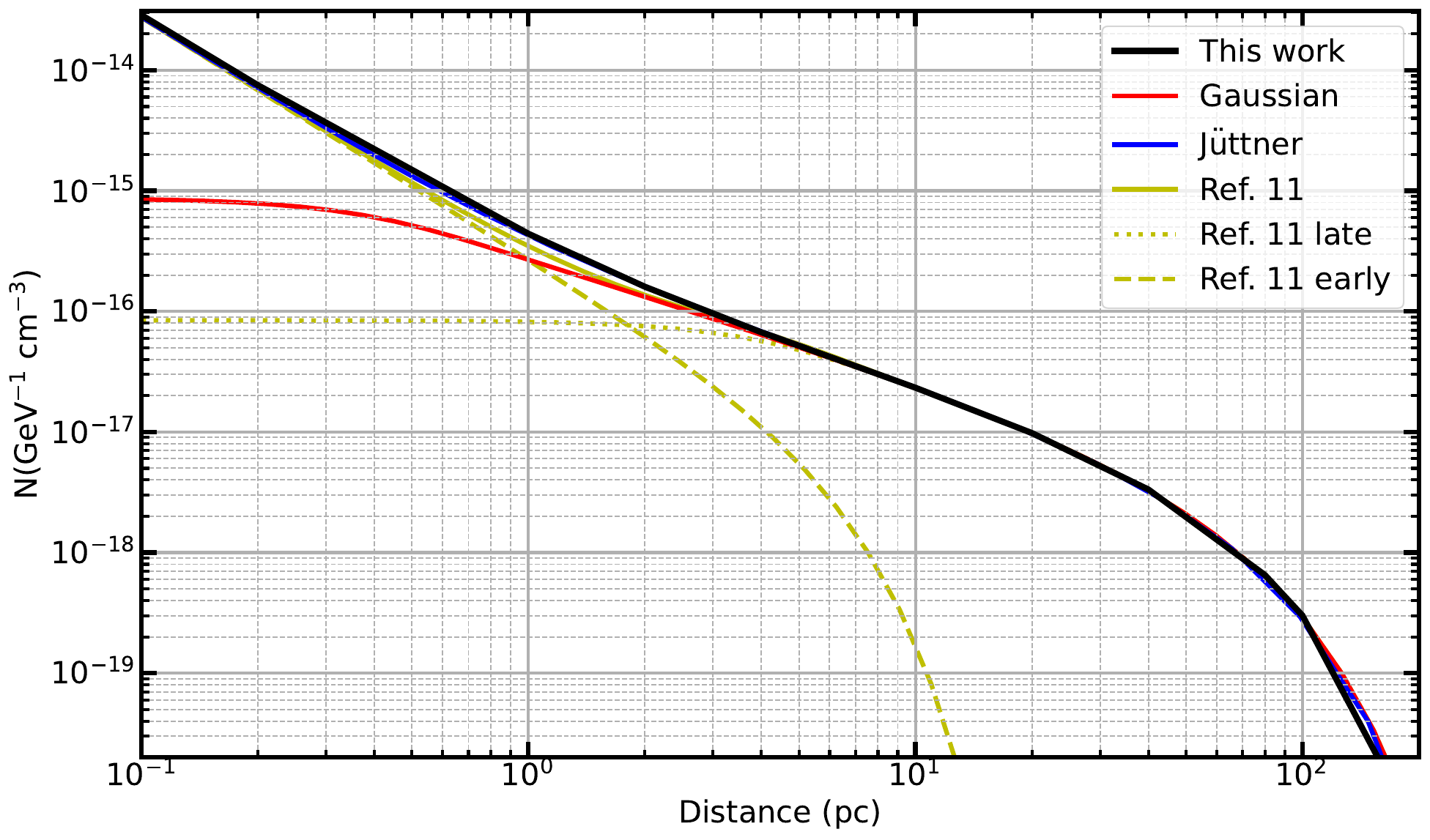}\\
\captionsetup{justification=raggedright}
\caption{Electron number densities as functions of the distance from the central pulsar for $D_0 = 2\times10^{27}\mathrm{cm}^2\mathrm{s}^{-1}$. The results derived using the propagator from this work, the Gaussian propagator, the generalized J\"{u}ttner propagator~\cite{Bao:2021hey}, and the result from Ref.~\cite{Recchia:2021kty} are shown as black, red, blue, and golden solid lines, respectively. The dotted and dashed golden lines correspond to the contributions from the ballistic and diffusive propagation in the calculation used by Ref.~\cite{Recchia:2021kty}.
\label{fig:profile}}
\end{figure}

\begin{figure*}[htbp]
    \begin{center}
        \subfloat{\includegraphics[width=0.45\textwidth]{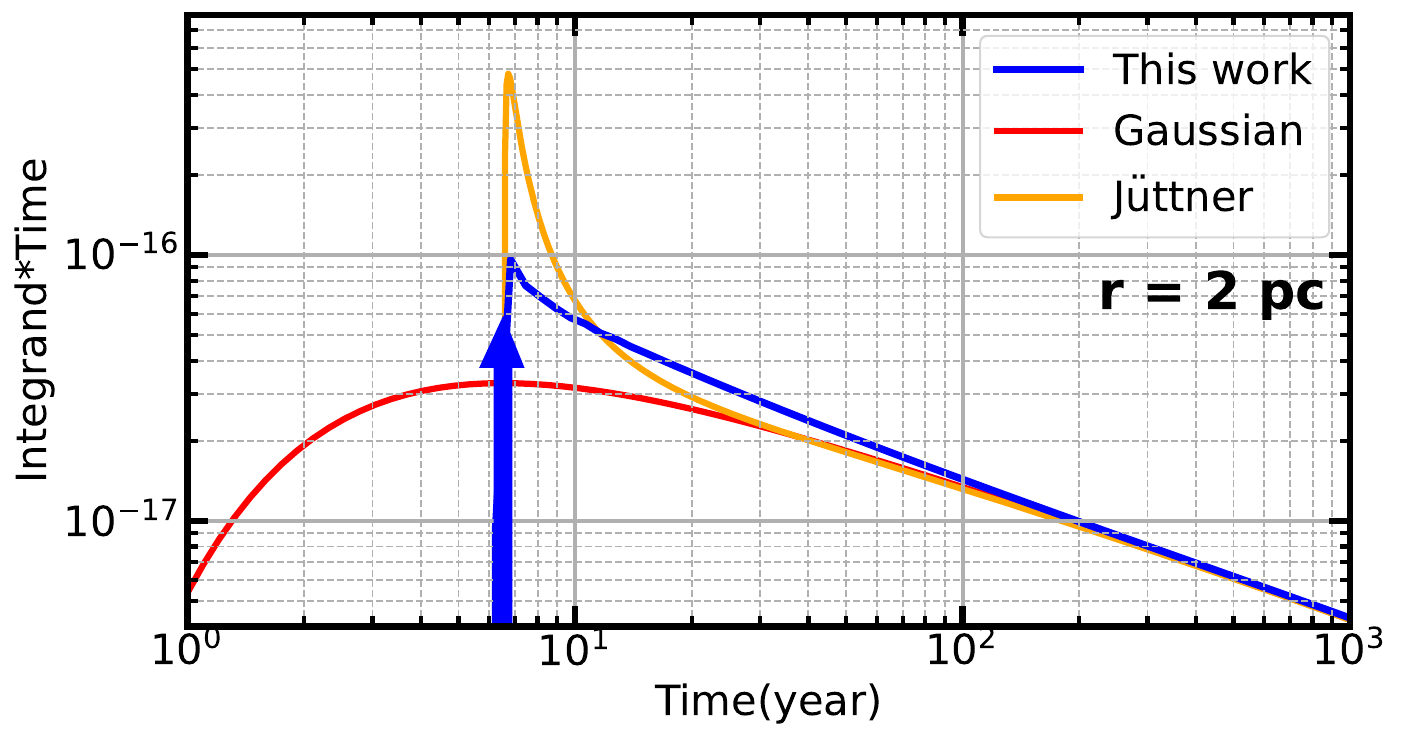}} \hskip 0.02\textwidth
        \subfloat{\includegraphics[width=0.45\textwidth]{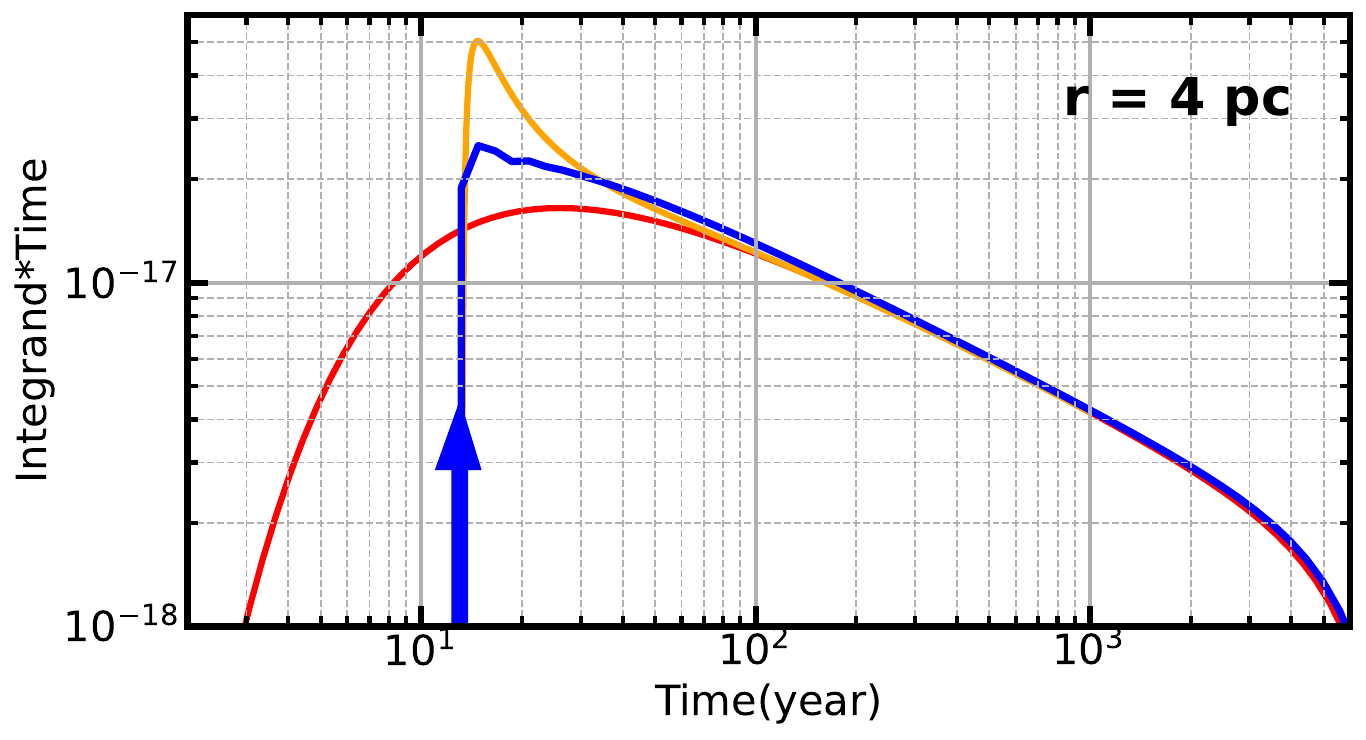}}\\
        \subfloat{\includegraphics[width=0.45\textwidth]{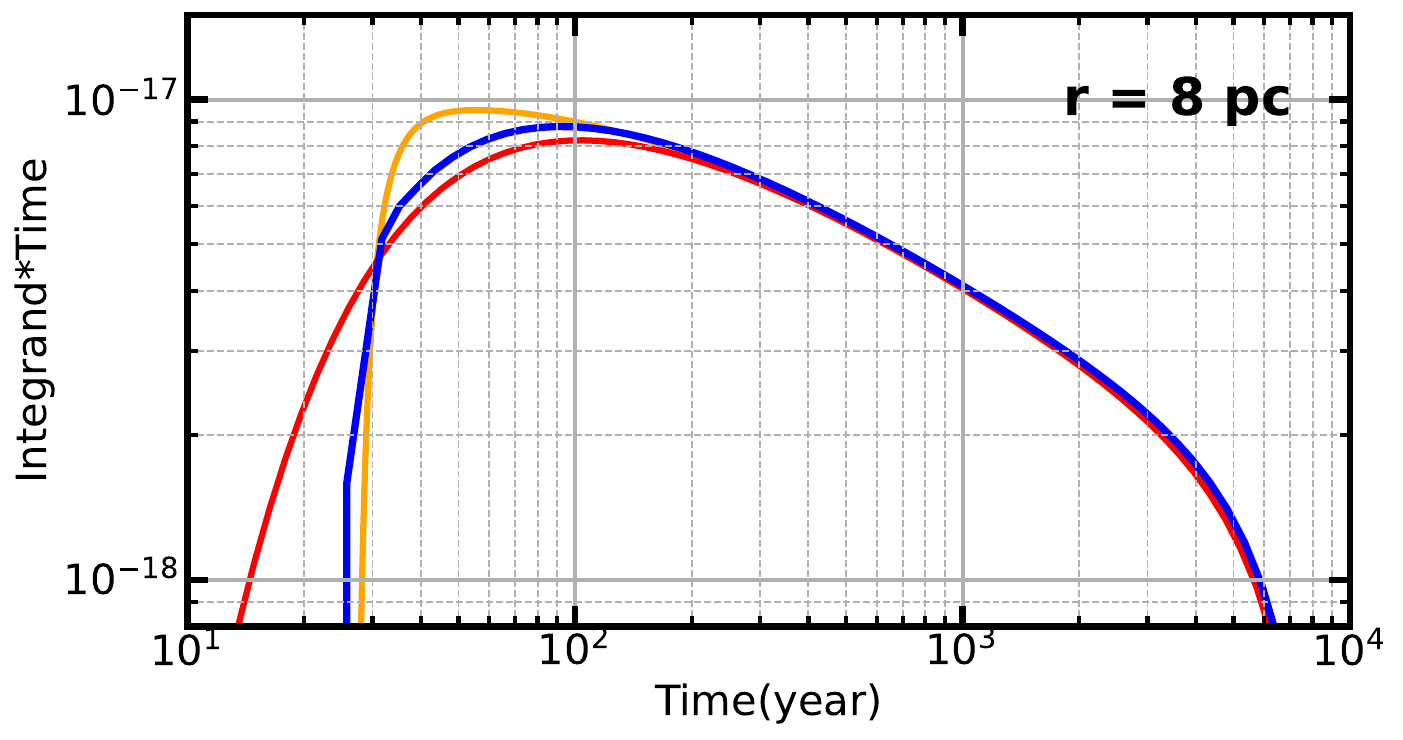}} \hskip 0.02\textwidth
        \subfloat{\includegraphics[width=0.45\textwidth]{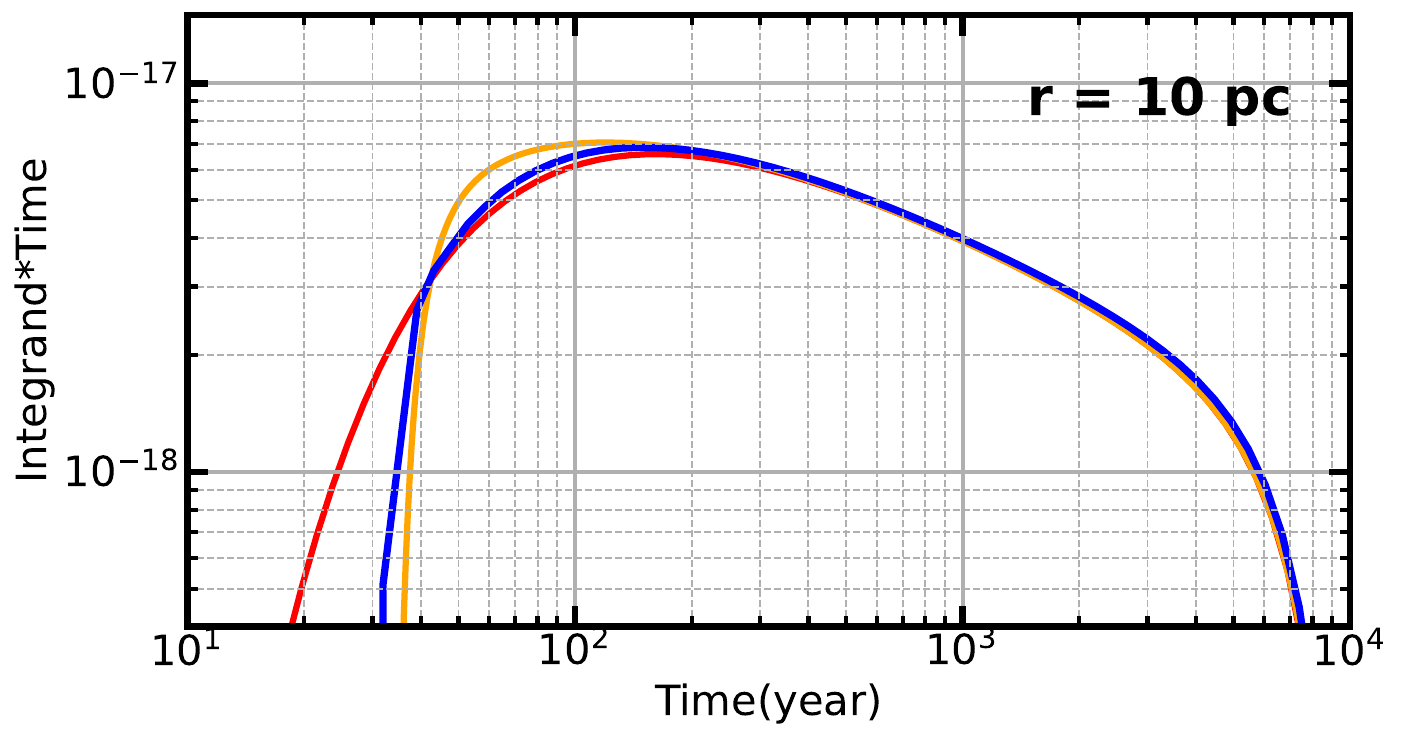}}
    \end{center}
    \captionsetup{justification=raggedright}
    \caption{Integrands of $e^\pm$ number density with respect to time for various distances from the central pulsar, as predicted by the propagator of this work, the Gaussian propagator, and the generalized J\"{u}ttner propagator, respectively. The arrows indicate contributions from ballistically moving particles without scattering.}\label{fig:integrand.}
\end{figure*}

It is evident that our approach shows good agreement with the simulation across different regimes: ballistic, diffusive, and the intermediate one. One might expect the heights of the arrows to match those of the last bin in the corresponding simulation. However, in the top row of Fig.~\ref{fig:green.}, the simulation results appear higher. This discrepancy arises because the last bin of the simulation includes contributions not only from non-scattered particles but also from scattered ones whose final positions are near the ballistic front. We have verified that, when only particles located precisely at the shock front are considered, our $\delta$-peak contributions agree well with the simulation results.

For the other two propagators, the Gaussian propagator clearly exhibits the superluminal diffusion problem, particularly for the time steps $t=5$ year and $t=$ 10 year, where it predicts that most particles are located well outside the future light cone of the particle source. The generalized J\"{u}ttner propagator is free from the superluminal problem due to the explicit introduction of the Heaviside step function in Eq.~(\ref{eqn: juttner prop}). However, it underestimates the mean free path of the diffusive process, which is evident across all six time steps. For instance, when \( t = 5 \) years, most particles are still moving ballistically since the mean time between scatterings is \( \tau = 10 \) years, meaning the majority of particles should be found in the \(\delta\)-peak. Although the generalized J\"{u}ttner propagator does produce a sharp peak that mimics this behavior, it still overestimates the fraction of particles that have undergone scattering. Moreover, as the time step increases, the Gaussian propagator converges to the simulation results more rapidly than the generalized J\"{u}ttner propagator, which, due to underestimating the mean free paths, leads to an overconcentration of particles at smaller radii.

\section{\label{sec:tevhalo}Application to pulasr halos}
Pulasr halos are extended gamma-ray structures produced by electrons and positrons ($e^{\pm}$) that have escaped from pulsar wind nebulae and are freely diffusing through the interstellar medium. To calculate the gamma-ray profile, it is essential to first determine the $e^{\pm}$ profile, with the detailed calculation provided in Appendix~\ref{sec:halo cal}. The Green's function required for the computation of Eq.~(\ref{eqn: diff-loss}) is related to the propagator obtained in Section~\ref{sec:results} through the relation:
\begin{equation}
\begin{aligned}
       G\left(E, \boldsymbol{r}, t ; E_0, \boldsymbol{r}_0, t_0\right) &= \rho(t-t_0, |\boldsymbol{r}-\boldsymbol{r}_0|;D_0) \\
       &\times\delta\left(t-t_0-\tau\right) 
       H(\tau) 
\end{aligned}
\end{equation}
where $\lambda$ and $\tau$ are given by Eq.~(\ref{eqn: lambda}).

Another approach to calculate the $e^{\pm}$ profile is proposed by Ref.~\cite{Recchia:2021kty} where the authors bypassed the use of a propagator and directly proposed a form for the electron profile incorporating superluminal corrections. They utilized the fact that the timescale $\tau_c = 3D/c^2$ can be viewed as the transition point between ballistic and diffusive propagation and combined these two regimes with a smoothing function. Specifically, the electrons propagate ballistically for $t < \tau_c$, with the density given by:
\begin{equation}
\begin{aligned}
N_{\text {ball}}(E, r, t) & =\int_{t-\tau_c}^t \frac{q(E) L(t)}{4 \pi c^3\left(t-t_0\right)^2} \delta\left(\left(t-t_0\right)-\frac{r}{c}\right) d t_0 \\
& =\frac{q(E) L(t)}{4 \pi c r^2} H\left(\tau_c c-r\right)\;.
\end{aligned}
\end{equation}
After the transition, for $t > \tau_c$, the electrons propagate according to the diffusion equation, with their density described by Eq.~(\ref{eqn:green method}), where the upper limit for the time integration is taken as $t-\tau_c$. To ensure a smooth transition between the ballistic and diffusive regimes, Ref.~\cite{Recchia:2021kty} replaced the Heaviside function $H\left(\tau_c c - r\right)$ with a smoothing function $\exp[-(r/(6D/c))^2]$. Thus, the total electron density is given by the sum of the two components: $N_{\text{tot}}(E, r, t) = N_{\text{ball}}(E, r, t) + N_{\text{diff}}(E, r, t)$.

The resulting $e^{\pm}$ profiles as functions of the distance from the central pulsar using different approaches for a reference diffusion coeffecient of $D_0 = 2\times10^{27}\mathrm{cm}^2\mathrm{s}^{-1}$ are plotted in Fig.~\ref{fig:profile}, where the individual contributions from ballistic and diffusive propagation predicted by Ref.\cite{Recchia:2021kty} are also shown. The electron energy is 100 TeV, which is roughly the parent electron energy of the TeV pulsar halos.

All three approaches that account for the maximum velocity at which electrons can move are in good agreement. Notably, our semi-analytic approach and the generalized J\"{u}ttner propagator differ by less than 10\% across all radii. The appraoch used by Ref.~\cite{Recchia:2021kty} aligns well with the other two in the ballistic and diffusive limits, as anticipated, but it underestimates the electron density in the intermediate regime. In contrast, the standard Gaussian propagator underestimates the electron density at small radii, which is a consequence of newly injected particles propagating with superluminal velocities. 

The good agreement between our semi-analytical propagator and the generalized J\"{u}ttner propagator may seem surprising, given that these two propagators differ quite a bit, as already shown in Fig.~\ref{fig:green.}.  The explanation lies in the fact that calculating the $e^{\pm}$ profile from the propagator requires an integration over time. Although the integrands of the two propagators are different, as illustrated in Fig.~\ref{fig:integrand.}, the total area underneath them is in agreement. Therefore, despite their differences at individual time steps, the overall contributions yield similar results when integrated over time. It means that for the case of continuous injection, the $e^{\pm}$ profile obtained using the J\"{u}ttner propagator is sufficiently accurate.

\section{\label{sec:conclusion}conclusion}
Diffusion equations are intrinsically nonrelativistic, and superluminal velocities naturally arise within them. Despite nearly a century of efforts, the fundamental solution to this issue—the relativistic generalization of the diffusion equation—remains elusive.

The propagation of CRs in galactic magnetic fields is a diffusive process. However, due to the intrinsic difficulties of the superluminal diffusion problem, previous studies relied on educated guesses to to address this challenge. One commonly adopted approach is the use of the generalized J\"{u}ttner propagator. Nevertheless, such methods lack a rigorous derivation from first principles, leaving uncertainty about how accurately they capture the true dynamics of particle propagation.

In this work, we take a different approach by starting from the microscopic picture of a random flight process, where particles move at a constant velocity with random scatterings, to derive the PDF of particle propagation semi-analytically. This provides a more physically grounded solution to the superluminal diffusion problem. We find the generalized J\"{u}ttner propagator does not provide a satisfactory description of the actual propagator. However, for the calculation of pulasr halos, where electrons and positrons are injected continuously, the resulting $e^{\pm}$ profiles from our approach and the generalized J\"{u}ttner propagator are similar after time integration. 

In contrast, for charged particles injected over a short period of time (burst-like injection), the discrepancies between the two approaches become significant, indicating that the generalized J\"{u}ttner propagator is inadequate for modeling transient events. Our semi-analytical solution, based on the random flight process, provides a more accurate and physically consistent description of particle diffusion.

We noticed that during the preparation of our paper, Ref.~\cite{Kawanaka:2024qiu} was published on the preprint site arXiv. Their work also aims to provide a rigorous solution to the superluminal diffusion problem with a similar model setup, and their results are comparable to ours. However, their computational approach differs significantly from ours. Specifically, they calculate the propagator by individually considering $0$ to $n$ scatterings and summing them to obtain the total propagator. In contrast, our approach is recursive, dividing the process only into non-scattered and scattered cases.

\acknowledgments
This work is supported by 
%the National Key R\&D Program Grants of China under Grant No. 2022YFA1604802, 
the National Natural Science Foundation of China under the Grants No. 12105292, No. 12175248, and No. 12393853.

\bibliography{apssamp}
\appendix
\section{\label{sec:halo cal}Calculation of electron profile in pulasr halos}
We consider the propagation of electrons escaping from pulsar wind nebulae and diffusing through the interstellar medium, which leads to the formation of pulasr halos. The electron propagation is governed by the diffusion-cooling equation:
\begin{equation}
\begin{aligned}
\frac{\partial N(E, \boldsymbol{r}, t)}{\partial t}=&D(E) \Delta N(E, \boldsymbol{r}, t)\\
&+\frac{\partial[b(E) N(E, \boldsymbol{r}, t)]}{\partial E}+Q(E, \boldsymbol{r}, t)\;,\label{eqn: diff-loss}
\end{aligned}
\end{equation}
where $E$ denotes the electron energy, and $N$ represents the electron number density. The diffusion coefficient is modeled as $D(E) = D_0(E/\text{1 GeV})^\delta$, with $\delta = 1/3$ based on Komolgorov’s theory. The term $b(E) = -\text{d}E/\text{d}t$ corresponds to the energy loss rate due to synchrotron radiation and inverse Compton scattering, and is expressed as $b(E) = b_0 E^2$, where $b_0 = 2.3 \times 10^{-17} \text{GeV}^{-1}\text{s}^{-1}$.

The source term is taken as:
\begin{equation}
Q(E, \boldsymbol{r}, t) 
= \begin{cases}q(E) L(t)\delta(\boldsymbol{r}-\boldsymbol{r}_s), & t \geqslant 0 \\
0, & t<0\end{cases},
\end{equation}
where $\boldsymbol{r}_s=\boldsymbol{0}$ is the position of the pulsar, $L(t) = (1+t / t_{\mathrm{sd}})^{-2} /(1+t_s / t_{\mathrm{sd}})^{-2}$, with $t_s$ being the pulsar's age and $t_{\mathrm{sd}} = 10$ kyr the typical spin-down timescale of the pulsar. The electron injection spectrum, $q(E)$, is assumed to follow a power-law form with an exponential cutoff, characterized by a spectral index $p = 1.0$ and a cutoff energy $E_c = 133\text{TeV}$.
%\begin{equation}
%q(E)=q_0(E / 1 \mathrm{GeV})^{-p} \exp \left[-\left(E / %E_c\right)^2\right]\;,
%\end{equation}
%with parameters $p = 1.0$, $E_c = 133 , \text{TeV}$, and $q_0 = 1.7 \times 10^{32} , \text{GeV}^{-1} \text{s}^{-1}$.

Eqn.~\ref{eqn: diff-loss} can be solved with the Green’s function method:
\begin{equation}
\begin{aligned}
N(E, \boldsymbol{r}, t)=&\int_{\mathbb{R}^3} \mathrm{~d}^3 \boldsymbol{r}_0 \int_{-\infty}^t \mathrm{~d} t_0 \int_{-\infty}^{+\infty}  \\
&\times\mathrm{d} E_0G\left(E, \boldsymbol{r}, t ; E_0, \boldsymbol{r}_0, t_0\right) Q\left(E_0, \boldsymbol{r}_0, t_0\right).
\end{aligned}\label{eqn:green method}
\end{equation}

And some quantites need for the calcualtion of $G\left(E, \boldsymbol{r}, t ; E_0, \boldsymbol{r}_0, t_0\right)$:
\begin{equation}
\lambda=2 \sqrt{\int_E^{E_0} \frac{D\left(E^{\prime}\right)}{b\left(E^{\prime}\right)} d E^{\prime}}, \quad \tau=\int_E^{E_0} \frac{d E^{\prime}}{b\left(E^{\prime}\right)}\;.\label{eqn: lambda}
\end{equation}
\section{\label{sec:eqns}Exact solutions of Persistent Random Walk in one, two, and four dimensions}
We list the solutions for 1,2, and 4 dimensions:
\begin{widetext}
\begin{equation}
\begin{aligned}
    \rho_1(x,t) =& e^{-\lambda t} \frac{\delta(x+vt)+\delta(x-vt)}{2} \\
                &+e^{-\lambda t}\frac{\lambda}{2 v}\left(I_0(\lambda\sqrt{t^2-(x/v)^2} + t\frac{I_1(\lambda\sqrt{t^2-(x/v)^2}}{\sqrt{t^2-(x/v)^2}} \right) H(vt-x)\;,
\end{aligned}
\end{equation}
\end{widetext}
where $I_{\nu}$ is the the modified Bessel function of the first kind of order $\nu$, and $H$ is the Heaviside step function.
\begin{widetext}
\begin{equation}
\rho_2(\vec{x}, t)=e^{-\lambda t}\frac{\delta(r-v t)}{2 \pi r}
+\frac{\lambda e^{-\lambda t}}{2 \pi v \sqrt{v^2 t^2-r^2}} \exp \left(\frac{\lambda \sqrt{v^2 t^2-r^2}}{v}\right) H(v t-r)\;,
\end{equation}
\end{widetext}
and
\begin{widetext}
\begin{equation}
    \rho_4(\vec{x}, t) = e^{-\lambda t}\frac{\delta(r-v t)}{2 \pi^2 r^3} + \frac{1}{\pi^2 v^4 t^3 \tau} \exp \left(-\frac{r^2}{v^2 t \tau}\right)\left[2+\frac{v^2 t^2-r^2}{v^2 t \tau}\right]H(v t-r)\;.
\end{equation}
\end{widetext}

%\nocite{*}
%
\end{document}